# Electromagnetic Radiation Reduction in 5G Networks and Beyond using Thermal Radiation Mode

Haneet Kour, *Student Member, IEEE*, Rakesh Kumar Jha, Senior *Member, IEEE*

*Abstract*— **With the massive increase in the popularity of smartphones and mobile data applications demanding bandwidth requiring data rates of the order of Gb/s, exploration of untapped frequency spectrum such as mmWave has begun. Along with providing seamless connectivity and catering to achieving high QoS and QoE, investigations are ongoing to enhance our knowledge about the biological safety at high frequencies. There is a need to ensure safety and reliability for the exposed public and updating the government policies regarding safety standards and regulations. This article is consecrated to provide an insight into health effects pertaining to mmWave frequencies, addressing aspects such as thermal heating in the body tissues with temperature rise, specific absorption rate (SAR), power density. As a solution a proposal has been given for EM (Electromagnetic) radiation reduction for the mobile communication system in the form of a proposed mode i.e. "Thermal Radiation" (TR) mode endorsing its safe use, promoting Green WCN along with increased energy efficiency and reduced complexity for the future generations to come. The proposal also validates reduced power density, SAR and temperature elevation produced in the human tissue when compared to other models in the form of simulation results obtained. It can increase the safety and reliability of 5G and beyond i.e. 6G networks in future.**

## I. INTRODUCTION

A remarkably high growth has been observed during the last decade in the wireless and mobile communication system since their inception in the 1970s [1]. Growth in the count of base stations that are being sited has severe implications due to EM radiations on humans, plant life, animals and the entire ecosystem at large [2]. There have been growing concerns and investigations regarding the adverse impact that the radiation exposure has caused on the biological safety of the beings [3-4]. Along with increasing the base station siting, research is initiated in expanding the spectrum band under use i.e. exploring the mmWave frequency band and so on.

H. Kour, R. K. Jha, are with the Shri Mata Vaishno Devi University, J&K, India. (E-mail: , hani.kpds@gmail.com, jharakesh.45@gmail.com,).

The EM radiation exposure limits and safety standards have been defined by the International Commission on Non-Ionising Radiation Protection (ICNIRP) in [5] regarding the permissible exposure up till 300 GHz exposure. Various reports by the National Radiological Protection Board (NRPB) of UK have also presented their guidelines for the same in [6], stating effects such as injury, illness, thermal heating of the human tissues. These concerns have also been highlighted in the recent events during ITU (International Telecommunication Union) [7].

The physical quantities used to study the exposure are specific absorption rate, Power Density (PD), electric and magnetic field. Wireless communication networks (WCN) have major contributions in increasing the carbon dioxide footprint in the atmosphere. Mobile networks alone contribute to more than half of the emissions followed by telecom devices. This figure is expected to rise even more in the near future [8]. Table I presents the EM radiation emission sources along with the frequency of operation and their transmission power. The adverse impacts from the same are manifold. In humans the eyes, skin tissues, brain are the most affected areas as they are under maximum exposure [9]. Radiation absorption by the eyes leads to corneal damage, the effects on skin can be witnessed from radiation absorption by the surface layer i.e. epidermis and by the underlying layer dermis. This leads to heating of the tissues due to temperature rise and it has also been validated that at high frequency of operation such as mmWave the power density exposure to the human body is also high [10].

Radiation exposure studies have also been made on animals such as mice and it has been found to cause a change in their gene expression, locomotion and reproductive health. Radiation poisoning has been found to occur in plants due to electrosmog. The emission of green house gases depletes the ozone layer cover and leads to extreme weather conditions which are on the rise every year. A pictorial representation of this has been presented in Fig. 1 depicting three layers. The innermost containing the cause, second depicting the impact and the outermost depicting the defense techniques to combat the EM radiation exposure. The radiation reduction techniques include device-to-device communication [11], spectrum sharing [12], Massive MIMO (Multiple Input Multiple Output) [13-14], coordinated multipoint [15-16] and RF shielding/Electromagnetic shielding for buildings or smart phones utilizing materials that reduce the specific absorption rates in human tissues [17]. The following section presents the related work in the literature highlighting the emerging concern and remedial techniques.

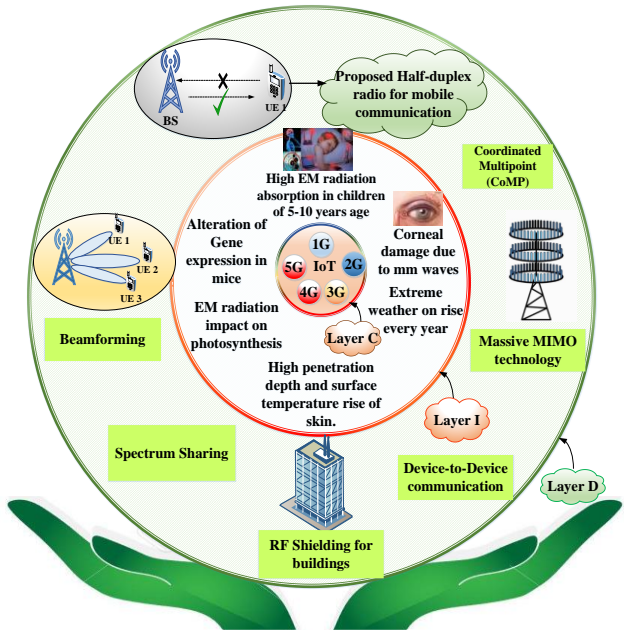

Fig. 1. Overall impact of EM radiation depicting cause, impact and defense techniques.

TABLE I
ELECTROMAGNETIC RADIATION EMISSION SOURCES.

| S.No. | EMF Source | Frequency | Transmission Power |
|---|---|---|---|
| 1. | AM/FM Radio | 540 KHz-1.6 MHz | 1 KW-30 KW |
| 2. | TV Tower | 48 MHz- 814 MHz | 10-500 W |
| 3. | Wi-Fi | 2.4 GHz | 10-100 mW |
| 4. | Cell Tower/ Base Station | 800/900/2100/2300 MHz | 20-100 W |
| 5. | Mobile Device | GSM- 900/1800MHz | 1W-2W |
| 6. | Radar | 1-100 GHz | 250 kW |
| 7. | mm Wave | Till 300 GHz | ~ 10 W |

## A. Related Work

Owing to the unprecedented traffic rush in the upcoming generation of wireless communications, several technologies/architectures have gained significant attention to increase the overall coverage, throughput in the network and improve the QoE of the users. The role of 5G NR (New Radio) has increased manifold as the future generation networks are required to be safe and not only fulfill the target data rate requirements. There are concerns regarding planning of the 5G networks in a way as to abide by the EMF limits and constraints [18]. Various 5G enabling technologies have been discussed in literature to provide new opportunities for reducing the exposure and for EMF aware network planning to plan the future communication networks [19]. In [20], the authors discuss how Massive MIMO systems can be incorporated in future generation architectures to increase the diversity gain and spatial multiplexing in the system. Since the total transmission power of the base station remains the same with only the number of antennas increasing, the power density levels automatically reduce hence decreasing the EM radiation exposure.

Coordinated multi-point [21], is another technique in which there is coordination among neighboring base stations so that the uplink and the downlink signals do not pose high interference problems in the network, hence increasing the overall spectral efficiency and throughput of the system. Other techniques that are gaining popularity with regard to enhancing the spectral efficiency of the system are spectrum sharing [12] and device-to-device communication (D2D) [11]. Spectrum sharing is an extremely efficient method for effective utilization of the available spectrum as there is utilization of T.V white spaces and underutilized part of the spectrum by secondary users. Device-to-device communication is also an efficient spectrum management method and helps in regulating the interference in the network using centralized or distributed power allocation schemes. Both spectrum sharing and D2D communication in the network can help in lowering the power density as well as EM radiation levels.

For the mobile communication systems, the radiations emitted from the cell phone are responsible for the harmful impact produced on the human body. RF (Radio Frequency) shielding/SAR shielding, involves coating with a ferrite material/conductive material between the handset and human head so that the specific absorption levels in the body are reduced. Studies have validated that using metamaterials can reduce the SAR by about 27%-52% at operating frequencies of 900 MHz and 1800 MHz [22]. Previously the base stations used in cellular communication had omnidirectional antennas which evolved to sectorized antennas. Sectorized antennas restricted the radiation power in a particular sector improving coverage but still produced inter-cell interference and also high radiation in the network.

Beamforming is a technique that involves antennas forming an array to focus the transmit beam in a desired direction towards a particular receiver [23-24]. This enhances the efficiency thereby reducing the transmit power and limiting the unwanted radiation in the network caused by surrounding users. Another method to limit the energy consumption by the base station is the dynamic "switching off" of the base station for heterogeneous networks. There are various BS (Base Station) switch off strategies [25] that have been studied in literature to encourage green communication in cellular networks hence making them safe and energy efficient. Antenna topologies such as adaptive array have been investigated in literature to mitigate the SAR for human head [26]. The antenna patterns are adjusted so as to improve the SINR (Signal-to-Interference-plus-Noise) and absorbed EM radiation in the head from mobiles.

All the existing methods and techniques help in EM reduction by either limiting the SAR or the power density in the network. The proposed Thermal Radiation (TR) mode is an effective and less complex solution to limit the EM radiation in the network. This is an effective approach that helps in limiting both the EM radiation metrics i.e. power density as well as SAR. Incorporating TR mode also reduces the interference and complexity in the network and hence can be an effective solution to move towards safe and green future generation networks. A comparison of some of the popular EM radiation reduction techniques with TR mode has been provided in Table II.

TABLE II
COMPARISON OF TR MODE WITH EXISTING EM RADIATION REDUCTION TECHNIQUES FOR WIRELESS COMMUNICATION

| S.No. | Reference | Objective | Reduction Technique Used | SAR Reduction | Mode of Operation (AM/TR)* | Power Density Reduction | Complexity |
|-------|-----------|-----------|--------------------------|---------------|----------------------------|-------------------------|------------|
| 1. | [13-14] | To reduce total transmit power of the base station. | Massive MIMO | ✗ | AM | ✓ | High |
| 2. | [15-16], [21] | Simultaneous communication between base stations to reduce interference and EM radiation. | Coordinated Multipoint (CoMP) | ✗ | AM | ✓ | Medium |
| 3. | [12] | Sharing the unused spectrum in time, frequency, space not utilized by primary users. | Spectrum Sharing | ✗ | AM | ✓ | Medium |
| 4. | [11] | Direct communication between devices to increase spectral efficiency. | Device to Device Communication | ✗ | AM | ✓ | Low |
| 5. | [22] | To block EM radiation field between mobile device and human head. | RF Shielding | ✓ (27-52%) | AM | ✗ | Medium |
| 6. | [23-24] | Direct the transmission beam towards desired receiver to reduce EM radiation towards surrounding devices. | Beamforming | ✗ | AM | ✓ | Low |
| 7. | [25] | Reduce the transmission power of the base station in the UL/DL decreasing interference and EM radiation exposure | BS Switch off strategy | ✗ | AM | ✓ | Low |
| 8. | [26] | Reduce the specific absorption levels of EM radiation for the human head and body. | Mobile phone Antenna Topology | ✓ | AM | ✗ | High |
| 9. | Thermal Radiation (TR) Mode | To limit the "always-on" transmissions reducing the RF energy and complexity. | Radiation Reduction technique | ✓ | TR | ✓ | Very Low |

*AM= Active Mode, TR= Thermal Radiation mode.

## B. Contributions and Organization:

In this paper, a Thermal Radiation (TR) mode has been proposed for mobile communication in WCN that considerably reduces the EM radiation exposure, with improved SINR and reduced complexity in the network. The major contributions are as follows:

- The article addresses a growing area of concern and investigation for the industry and academia i.e. ensuring that the expansion in the wireless and mobile communication is made safe and reliable in operation for the future generations by reducing the RF exposure in the network.

- Mobile communication systems have been particularly addressed in the article and a proposal in the form of "Thermal Radiation" (TR) mode has been given for the smartphone that reduces the radiated emissions hence reducing the power density, SAR levels in the human head and other body tissues.

- The simulation results obtained with the proposal validate that with the introduction of Thermal Radiation mode in the mobile handset, the thermal heating of the tissues decreases and hence the temperature elevation produced in them.

- Introduction of TR mode also helps in making the network more energy efficient hence contributing towards Green communication.

- The proposal helps in mitigating the overall interference in the network which has been validated in the form of

reduced complexity obtained for the proposal in comparison to the current communication scenario.

The rest of the article is organized as follows. System model and problem formulation have been presented in Section II. The next section III depicts realization and representation of Thermal Radiation (TR) mode. Section IV illustratively explains TR mode considering a practical example. The simulation results and discussions have been given in section V followed by comparative analysis in Section VI. The article concludes in Section VII.

## II. SYSTEM MODEL AND PROBLEM FORMULATION

The system model for analyzing the proposed thermal radiation mode for future mobile communication systems is presented in this section, followed by problem formulation.

### A. System Model

The overall architecture consisting of 5G Core (5GC) network and NG-RAN (Next-Generation Radio Access Network) is depicted in the system model (Fig.3). 5GC consists of AMF/UPF (Access and Mobility Management Function/ User Plane Function) which have their respective main functionalities. AMF is responsible for NAS (Non-Access Stratum) signaling termination/signaling security, access stratum security control. It also supports intra and inter system mobility along with providing access authentication/authorization [27]. UPF aids in routing and forwarding of the packet. It serves as an anchor point wherever there is Intra/Inter RAT mobility applicable. UPF also handles verification of uplink traffic and buffering of downlink packet.

NG-RAN consists of NG-RAN nodes which is either a gNB or an ng-eNB. A gNB provides the protocol termination of New Radio user plane and control plane towards UE. An ng-eNB provides the protocol termination of E-UTRA user and control plane towards UE. Both the NG-RAN nodes are interconnected through Xn interfaces. The NG-RAN nodes (gNB and ng-eNB) are also connected to AMF by NG-C interface and to UPF by NG-U interface.

Further we consider ng- eNB connected to user equipments (UE's) depicting two different cells. The UE's are denoted as $(UE_1, UE_2, UE_3, \ldots \ldots UE_{50})$ in both the cells assuming 50 UE's in each cell. One of the cell consists of all the users communicating in Active Mode (always-on links) and the other cell consists of some users in TR mode and the remaining in Active Mode. The cells are depicted by "Active mode zone" and "Thermal mode zone" respectively. In the "Thermal mode zone" we assume 30 users in AM mode and 20 users in TR mode.

The channel statistics are assumed as Non-stationary and wideband in nature considering Rayleigh fading distribution. Since the received Signal-to-Noise ratio fluctuates rapidly due to multipath, reflections, fading and so on, a non-stationary channel assumption is made. Wideband channels are considered to accommodate large bandwidth requiring applications for 5G and beyond 5G communication scenario. Wideband communication channels are suited for high bandwidth applications with large data capacities, because continuous data, voice and video applications have to be transmitted and received.

*Proof*: The proof is given in Appendix A, **Lemma 1**. ∎

We consider a 5G NR transmission frame $T_f$ of 10ms duration. The uplink transmissions take place by the user equipments to the base station utilizing the allocated orthogonal subcarriers. The frame is divided into subframes and further into '$M$' slots. The slots are designated as $(\tau_1, \tau_2, \tau_3, \tau_4, \ldots \ldots \tau_M)$. For a particular time slot the total power allocated by the base station is depicted as $p_i^t = (p_{i,1}^t, p_{i,2}^t, p_{i,3}^t, \ldots. p_{i,K}^t)$ to '$K$' user equipments over the downlink orthogonal subcarriers using OFDMA (Orthogonal Frequency Division Multiple Access) scheme. The maximum power limit for each time slot is denoted as $P_{max}$.

Every transmission frame $T_f$ has an average power constraint $P_{avg}$ for each frame. The above two constraints can be depicted as

$$\sum_{j=1}^{K} p_{i,j}^t \leq P_{max}, \quad i = 1,2 \ldots \ldots M \tag{1}$$

$$\sum_{i=1}^{M} \sum_{j=1}^{K} p_{i,j}^t \leq MP_{avg} \tag{2}$$

Considering a specific user equipment, Fig.2 depicts the functional module for a user $U_j$. It consists of the decision device which is responsible for deciding the mode in which the user equipment is operating. For a particular time slot $\tau_i$, the power splitter is responsible for splitting the power into two different portions one of which flows into the TR mode circuitry and the remaining part for the active mode where information decoding and processing takes place.

If the UE is in airplane mode/flight mode there is no requirement of information decoding or processing required for it.

The power splitting ratio for the particular time slot $\tau_i$ is denoted by $\alpha_{i,j} \in [0,1]$. The portion of the power dedicated for the TR mode is denoted as $p_{i,j}^{TR}(\alpha_{i,j}, p_i^t)$ and that for active mode (AM) as $p_{i,j}^{AM}(\alpha_{i,j}, p_i^t)$. The expressions for the same are articulated as follows:

$$p_{i,j}^{TR}(\alpha_{i,j}, p_i^t) = \gamma\, h_{i,j}(1 - \alpha_{i,j}) p_i^{DL} \tag{3}$$

$$p_{i,j}^{AM}(\alpha_{i,j}, p_i^t) = \alpha_{i,j}\, h_{i,j}\, (p_i^t - p_i^{DL}) \tag{4}$$

In the above equations (3) and (4) '$\gamma$' is the rate of conversion of RF signals to direct current (DC) and is assumed to be unity for the sake of simplicity. $h_{i,j}$ is the channel power gain of the assigned subcarriers for the downlink transmission for slot $\tau_i$. It is assumed that the user equipment utilizes the energy harvested from the TR mode for the uplink transmission of data.

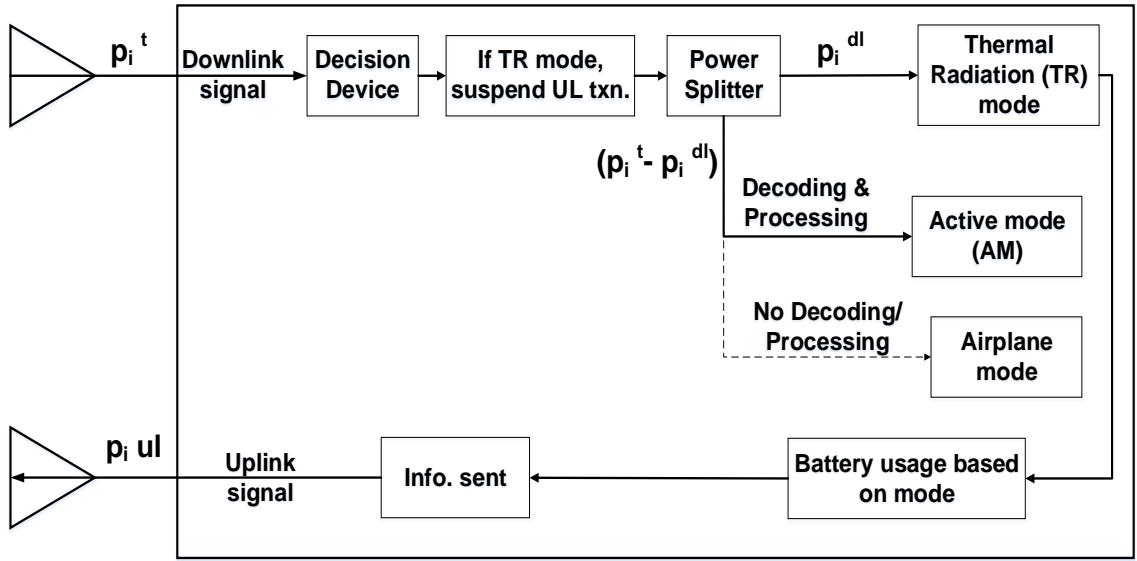

Fig.2 Functional module of a user equipment.

$p_i^{DL}$ is the portion of the total power allocated to be used for the TR mode and $(p_i^t - p_i^{DL}) = p_i^{UL}$ is the remaining portion of the total power for decoding and processing the information when the user equipment is operating in the active mode. The amount of power saved when the UE is in TR mode is more in comparison to when it is operating in the active mode.

Let us assume a scenario consisting of a base station and 50 user equipments (UE) operating in a cellular communication network. We intend to analyze the power utilized when all the UE's follow the conventional Active Mode communication v/s when some of the UE's are in TR mode and remaining in active mode. The total power utilization (for both downlink and uplink transmission) when all 50 users are in active mode can be represented as

$$P_{AM} = \sum_{j=1}^{50} [\, p_i^{DL}((1-\alpha_{i,j}), p_i^t) + p_i(\alpha_{i,j}, (p_i^t - p_i^{DL}))\,] \quad (5)$$

Similarly, if we assume 30 users are operating in the active mode and remaining 20 in TR mode, then the total power utilization is formulated as

$$P_{AM+TR} = \sum_{j=1}^{30} [\, p_i^{DL}((1-\alpha_{i,j}), p_i^t) + p_i(\alpha_{i,j}, (p_i^t - p_i^{DL}))\,] + \sum_{j=31}^{50} [\gamma\, p_i^{DL} h_{i,j}\,(1-\alpha_{i,j})] \quad (6)$$

Total power saved when some users operate in TR mode can be represented as

$$P_{saved} = P_{AM} - P_{AM+TR} \quad (7)$$

Here $P_{saved}$ is the power conserved when 30 users are communicating in the active mode and remaining users 20 are communicating in the TR mode.

Resultantly the downlink and the uplink throughputs can be calculated for the UE when it is operating in the active mode as well as the downlink throughput rate when operating in the TR mode. The downlink and the uplink throughputs when the UE is operating in the active mode are depicted by $R_{i,j}^{DL}(\alpha_{i,j}, p_i^{DL})$ and $R_{i,j}^{UL}(\alpha_{i,j}, p_i^{UL})$ in bits/s/Hz respectively and formulated as follows:

$$R_{i,j}^{DL}(\alpha_{i,j}, p_i^{DL}) = \sum_{i=1}^{K} log_2\left(1 + \frac{p_{i,j}^{DL}(\alpha_{i,j}, p_i^t)}{\sigma_n^2}\right), \quad (8)$$

$$R_{i,j}^{UL}(\alpha_{i,j}, p_i^{UL}) = \sum_{i=1}^{K} log_2\left(1 + \frac{p_{i,j}^{UL}(\alpha_{i,j}, p_i^t - p_i^{DL}) h_{i,j}^{UL}}{\sigma_n^2}\right), \quad (9)$$

where $h_{i,j}^{UL}$ is the power gain of the uplink channel in the transmission slot $\tau_i$. The throughput obtained when the UE is operating in the TR mode is depicted as $R_{i,j}^{TR}(\alpha_{i,j}, p_i^{DL})$ in bits/s/Hz and formulated as:

$$R_{i,j}^{TR}(\alpha_{i,j}, p_i^{DL}) = \sum_{i=1}^{K} log_2\left(1 + \frac{p_{i,j}^{DL}(\alpha_{i,j}, p_i^t)}{\sigma_n^2}\right). \quad (10)$$

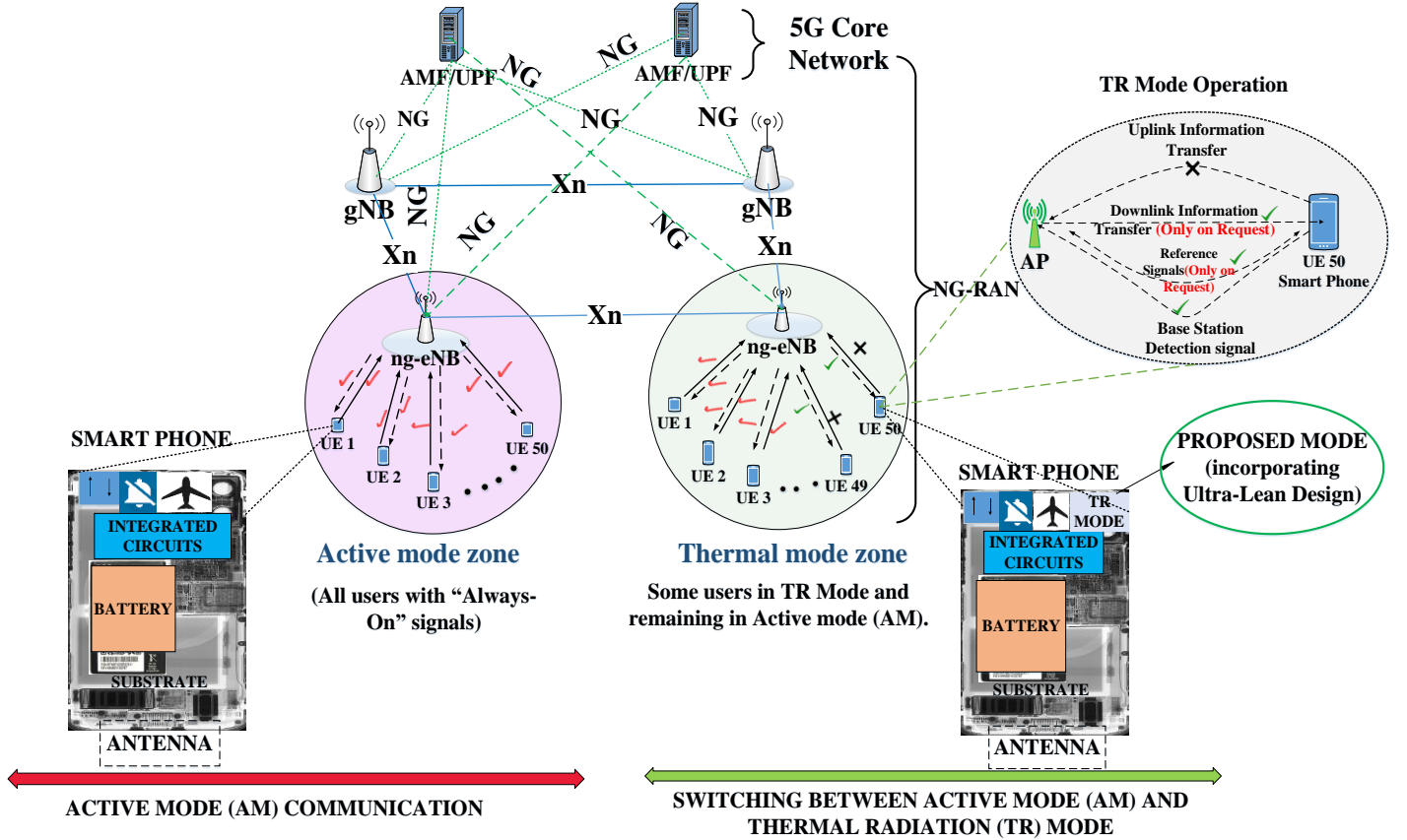

Fig. 3 System Model.

**Proposition 1:** The UE's communicating in the TR mode utilize a small portion of the total power as the uplink transmissions are suspended. The specific absorption rate (SAR) in case of UE's in TR mode reduces as the body tissue is now exposed to low radiation power. SAR for the device in TR mode can be formulated as

$$SAR_{TR} = \frac{p_i^{DL}}{M},\qquad(11)$$

where $p_i^{DL}$ is the radiation power to which the tissue is exposed for a device in TR mode and $M$ is the mass of the tissue in Kg. The value of $SAR_{TR}$ is lower than that of a device communicating in the Active (AM) mode, as we have both uplink and downlink powers contributing to the total SAR encountered i.e.

$$SAR_{AM} = \frac{(p_i^t - p_i^{DL})}{M}.\qquad(12)$$

in case of AM mode which is a much higher value than $SAR_{TR}$ as the former consists of both uplink and downlink powers contributing to the total SAR.

**Proposition 2:** For the cell having all the active users communicating in Active mode, both uplink and downlink signals contribute to the power density levels in the atmosphere. The devices in TR mode only have downlink signals active and the uplink transmissions are suspended. Hence, the far-field radiation density is lower in the cell where some users are operating in the TR mode. The Power Density (PD) radiated at '$d$' distance from the radiating antenna can be expressed as

$$PD = \frac{G_{tr}\,P_T}{4\pi d^2}\qquad(13)$$

where $G_{tr}$ is the gain of the transmit antenna, $P_T$ is the total input power to the transmit antenna and '$d$' is the distance from the radiating source. In TR mode the power density ($P_{TR}$) to which the body tissue is exposed is lower than the power density value for the cell consisting of all users in Active mode ($P_{AM}$).

### B. Problem Formulation

The proposal aims to minimize the EM radiation exposure by reducing the power density levels, specific absorption rate and the temperature elevation produced in the tissues due to high SAR maintaining high quality of service and experience of the users. We also aim to maximize SINR and reduce the overall complexity of the network.

The optimization problem can be formulated as follows:

$$\text{Max } R_{i,j}^{TR}\big(\alpha_{i,j}, p_i^{DL}\big)\qquad(14)$$

s.t.

$$S(x)\ i.e\ \sum_{j=1}^{30} SAR_{AM} + \sum_{j=31}^{50} SAR_{TR} < \sum_{j=1}^{50} SAR_{AM}\ (14(a))$$

$$PD = \sum_{j=1}^{30} PD_{AM} + \sum_{j=31}^{50} PD_{TR} < \sum_{j=1}^{50} PD_{AM}\quad(14(b))$$

Constraint (14(a)) ensures that when all the considered users (here 50) are operating in Active mode then the specific absorption rate obtained collaboratively for all the active users is higher than that obtained for the cell where some users operate in Active mode (30) and remaining (20) in TR mode. Constraint (14(b)) ensures that the power density values when all the communicating users in the cell follow Active mode of communication is much higher than that obtained in the "thermal mode zone" where not all users are operating in AM mode. In the TR mode due to suspension of uplink signals the radiated power density reduces considerably when some users operate in Active mode and some in TR mode.

### III. REALISATION AND REPRESENTATION OF THERMAL RADIATION (TR) MODE

This section presents the flowcharts and state transition diagrams to gain an insight into the realization of Thermal radiation mode for mobile communication system. The flowchart in Fig. 4 depicts the transition of the mobile between the Active mode, Thermal Radiation mode and flight mode.

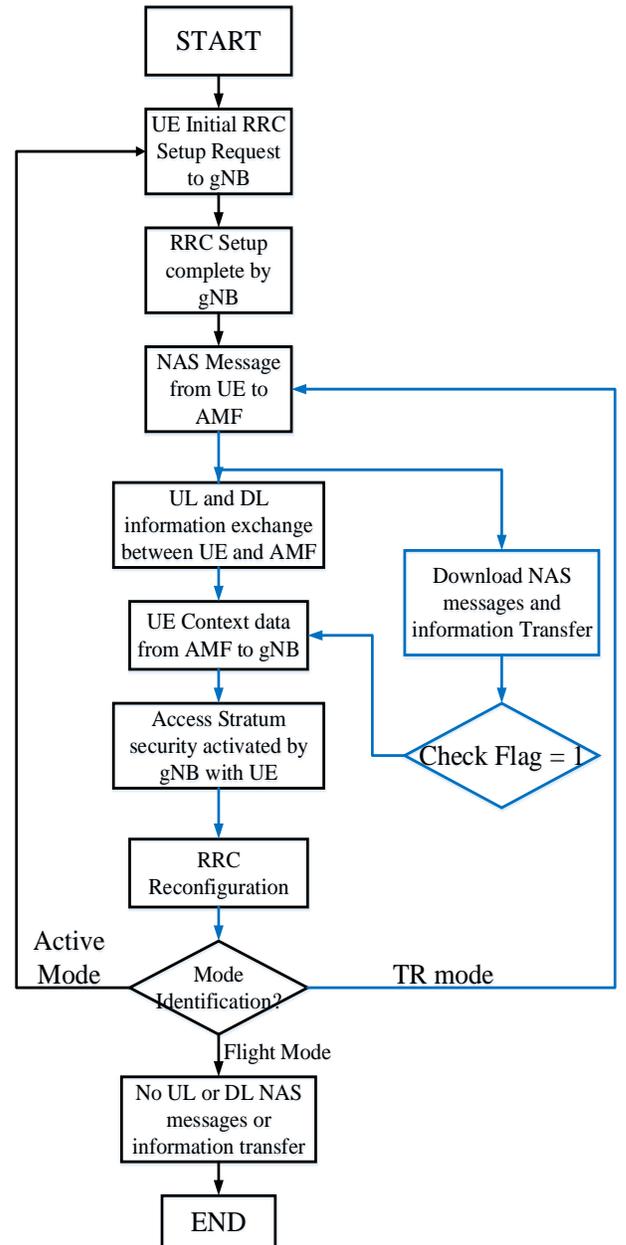

Fig. 4 Flowchart of UE state transition for Active mode, TR mode and flight mode.

Fig. 5 presents the user equipment triggered state transition diagram from idle mode to connected mode for TR mode. A potential state transition model has also been given for Active to TR mode in Fig. 6. The flowchart depicts the call setup procedure from Idle to connected state for Active, TR and flight mode. For all the three states there is difference in the transition steps involved. For the Active mode, the procedures involved are same as conventional as in [27]. For the TR mode, on completion of call setup i.e. the transition of state from RRC_IDLE to RRC_CONNECTED the User Equipment (UE) message from gNB to AMF initiates only the Downlink NAS transport and DL (Downlink) information transfer following it. The uplink information transfer and NAS transport are suspended in TR mode. The flag status is checked equal to 1 which confirms the activation of thermal radiation mode. The procedures following it are resumed as in Active mode.

On encountering the flight mode after mode identification, there is no UL or DL NAS message or information transfer between the UE, gNB and AMF. The proposed state transition for the user equipment from Idle to Connected state has been described in Fig. 5. The transition is user equipment triggered and is completed by gNB. After the RRC setup is complete and the access stratum (AS) security with the user equipment is activated by the gNB, there is reconfiguration performed by it. Following this there is mode identification i.e. whether active or in TR mode. If the TR mode is on, the flag status is updated to 1. This activates the TR mode of communication in the mobile handset. Only downlink transfer signals flow during the TR mode and the uplink transmissions are suspended until the transition state shifts from RRC EE to RRC Active mode.

A potential state transition model (Fig.6) is proposed for the TR mode with additional RRC states i.e. RRC_AM/TR and RRC_EE. The existing standardization for 5G NR consists of RRC_Idle, RRC_Inactive and RRC_Connected each exhibiting different characteristics. The proposed RRC_AM/TR mode is the transition mode before the phone transitions to TR mode. RRC_EE is a low activity node and it keeps the 5GC-NG-RAN connection ON only for downlink NAS and information transfer. This reduces the overhead signaling in the air interface for the TR mode and promotes energy saving for URLLC (Ultra-reliable low latency communication) services in 5G NR.

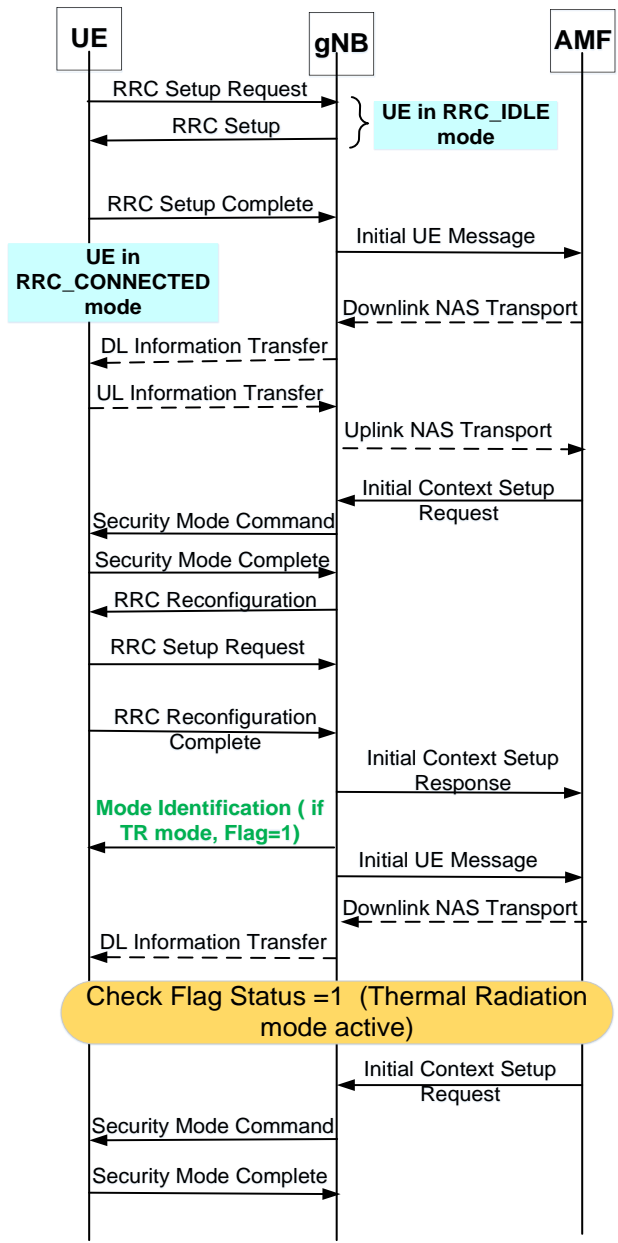

Fig.5. User Equipment triggered transition from IDLE to CONNECTED mode for TR mode.

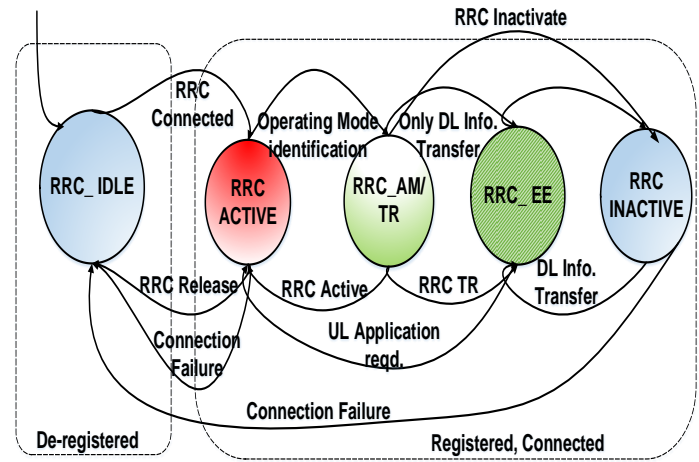

Fig.6 State transition model for Active to TR mode.

## IV. TR MODE: AN ILLUSTRATIVE EXAMPLE

The framework of the system for an illustrative example is presented in Fig.7 where in the current scenario we consider a mobile phone communicating in Active mode with the signal strength bars appearing at the top notification bar of the phone. We know that the received signal strength on the phone varies in accordance with the channel conditions between the transmitting base station and the receiving user equipment. These variations in the channel conditions sometimes help in achieving excellent signal strength corresponding to display of the signal strength as all active bars on the device. When the signal deteriorates due to varying channel conditions such as fading, scattering and non-line of sight communication, the number of active bars also decrease. In our system model we put forward the proposition that the mobile phone follows "display and decision technique $(D^2)$".

When the mobile phone receives great/good signal strength i.e. in the range of -50 dBm to -89 dBm due to line of sight communication or NLOS (Non-Line of Sight) constructive phenomenon, all the bars depicting the signal strength depict green color (Fig. 7). When the signal strength reduces i.e. the received signal strength is average in the range -90 dBm to -99 dBm due to NLOS communication, the signal strength bars go orange symbolizing average signal strength. If the received signal strength further deteriorates i.e. in the range of -100 dBm to -120 dBm, it becomes very difficult to manage both outgoing and incoming calls and the radiated electromagnetic intensity is also very high from the mobile phone. We propose that if the signal strength falls below the average strength, the mobile phone automatically transitions from the active mode to the TR mode i.e. the applications demanding high data rate services cannot be serviced in the TR mode until the signal strength improves.

***Assumption 1:*** *Some UEs are assumed to be operating in the Active mode and some in the TR mode. All the mobile phones are assumed to be operating with one omnidirectional antenna and demanding three different applications i.e.* $A_1 = Text$, $A_2 = Conversational\ Voice$, $A_3 = Conversational\ Video$.

### A. Channel Model

We consider that the UEs operating in Active mode as well as TR mode follow orthogonal frequency division multiplexing (OFDM) and experience Rayleigh fading. The active UEs are considered to be demanding applications, $Appl = [A_1, A_2, A_3.]$ First we consider the transmissions for the Active mode and then for the TR mode.

#### 1) Active Mode:

The received signal at a $i^{th}$ UE communicating in the active mode i.e. following Active Mode communication and transmitting signal $x_i^{AM}$ can be denoted as

$$y_i^{AM} = \sqrt{P_T^{AM}}\, h_{BS,MU}\, x_i^{AM} + Z^{AM}, \tag{15}$$

where $h_{BS,MU}$ represents the channel coefficient between the base station and the mobile user, $Z^{AM}$ is the noise vector at the receiver having zero mean and variance $\sigma_0^2$. The signal-to-noise (SNR) ratio received at the mobile user is denoted as

$$SNR_i^{AM} = \frac{P_T^i\, |h_{BS,MU}|^2}{\sigma_0^2}. \tag{16}$$

The SINR for every user communicating in Active Mode is given as follows

$$SINR_i^{AM} = \frac{P_T^i\, |h_{BS,MU}|^2}{\sigma_0^2 + I_{c(j)}}, \tag{17}$$

$I_{c(j)}$ is the total interference encountered by the cellular user which is given as the sum of multiple factors

$$I_{c(j)} = I_{BS} + I_{UE} \tag{18}$$

$$= T_{BS}\, h_{BS,MU} + P_j h_{j,j'} \tag{19}$$

where $I_{BS}$ is the interference encountered from the BS and $I_{UE}$ the interference from other UE's. $T_{BS}$ (Transmission power of the BS) and $P_j$ (transmission power of the UE) are assumed to be fixed. The interference due to the other cellular users in the cell is given by

$$I_{UE} = P_j h_{j,j'}. \tag{20}$$

The total interference to the users communicating in Active mode is a result of all the applications being Active i.e. $A_1, A_2, A_3$ and all the uplink and downlink signals ongoing which can be written as follows

$$I_{c(j)}^{A_1,A_2,A_3} = |h_{i,j}^{UL}|^2 + |h_{i,j}^{DL}|^2 \tag{21}$$

The interference power makes the system complex and reduces the achievable data rates in the network. Considering an available bandwidth of $\dot{B}$ Hz different achievable rates corresponding to different applications $A_1, A_2, A_3$ can be denoted as

$$RT_{1,2,3}^i = \dot{B}\, log_2(1 + SNR_i^{AM}) \tag{22}$$

We consider that there is optimum resource allocation to the UE's based on the demanded application and the optimum power for a required target rate is given as $PT_{opt}^{A_{1,2,3}}$, for a particular application.

Using the Shannon capacity theorem the target rate for a particular application $A_1$ is given as

$$RT_i^{A_1} = \dot{B}\, log_2(1 + \frac{P_{T(opt)}^{A_1}\, |h_{BS,MU}|^2}{\sigma_0^2}), \tag{23}$$

On rearranging the above equation we can calculate the optimum value of the transmission power

$$P_{T(opt)}^{A_{pppl}} = \frac{\sigma_0^2}{|h_{BS,MU}|^2}\left(2^{RT_i^{A_1}}/\dot{B} - 1\right) \tag{24}$$

Considering all the applications and the corresponding achievable rates as $R_1, R_1, R_2, R_3$, the total optimum power for serving all the applications when summed up can be denoted as

$$P_{T(max)}^{Total\ (AM)} = \frac{\sigma_0^2}{|h_{BS,MU}|^2}\Big\{\left(2^{RT_1^{A_1}}/\dot{B} - 1\right) + \left(2^{RT_2^{A_2}}/\dot{B} - 1\right) + 2^{RT_3^{A_3}}/\dot{B} - 1\Big\}. \tag{25}$$

When the signal strength falls below the average limit, we propose that the mobile phone transitions from active mode to TR mode.

#### 2) Thermal Radiation (TR) Mode:

In the TR mode the applications requiring very high bandwidth requirements will not be served. Optimum power will be required only to serve applications $A_1$ and $A_2$. Here the interference power in the network is only due to $A_1$ and $A_2$ as there is suspension of uplink information transfer for the applications with high data rate. The equation for interference power in TR mode can be reframed using eq. (21) as follows

$$I_{c(j)}^{A_1,A_2.} = |h_{i,j}^{UL}|^2 + |h_{i,j}^{DL}|^2, \tag{26}$$

The total optimum power required to serve the two applications with corresponding rates $R_1$ and $R_2$ can be summed up as follows

$$P_{T(opt)}^{Total\,(TR)} = \frac{\sigma_0^2}{|h_{BS,MU}|^2}\left\{\left(2^{RT_1^{A_1}}/\dot{B}-1\right) + \left(2^{RT_2^{A_2}}/\dot{B}-1\right)\right\}, \quad (27)$$

Considering a user in active mode and a user in TR mode, both being served the considered applications, the total power saved can be written as

$$P_{saved} = P_{T(max)}^{Total\,(AM)} - P_{T(opt)}^{Total\,(TR)}. \quad (28)$$

The power that is saved can be used to serve the applications when the signal strength is appropriate. Reduction in the utilization of power of all the users in TR mode reduces the SAR levels as the power dissipation from the phone reduces which leads to low absorption of harmful radiations into the head and body.

### B. Problem Formulation

We know that the next generation of WCN are required to achieve the high data rate requirements, optimize the power levels whilst maintaining the QoS and QoE and reducing the EM radiation levels. As the TR mode does not support high bandwidth requiring applications it is important to uphold a tradeoff between the achieved rates and the optimum power required.
We set a threshold limit for the signal strength, below which if the signal strength reduces, there is transition from the active mode to the TR mode. The threshold limit ($SS_{th}$) for the signal strength has been set to $SS_{th} = -99\,dBm$. The energy efficiency (EE) for a particular application is represented as

$$EE_i = \frac{RT^{appl}}{\sum P_T^{T(i)} + P_{ckt}} \quad (29)$$

The optimization problem is presented as

$$\max_{P_T^{T(i)}} EE_i \quad (30)$$

s.t. $\quad Signal\ Strength\ (SS) > -99\,dBm\ (SS_{th})\quad (30a)$

$$-99dBm < A_1, A_2, A_3 < -50\,dBm \quad (30b)$$

As $Signal\ Strength\ (SS) < SS_{th}$ transit to TR mode $(30c)$
The constraint (30(a)) states that the received signal strength in the mobile phone must be more than the threshold limit set for all the considered applications to be served to the UE. If the signal strength falls below the set threshold, then the applications requiring high bandwidth will not be served as it would lead to large emitted radiation from the mobile equipment and reduce the biological safety for the users. Constraint (30(b)) makes sure that for the considered applications to be served, the received signal strength must lie within the given range for controlled emission radiation. The constraint (30(c)) depicts that as the signal strength drops below the threshold limit the mode of the phone changes from active to TR, hence not allowing high bandwidth requiring applications to be served. Table III presents the values corresponding to different ranges for three different signal operators.

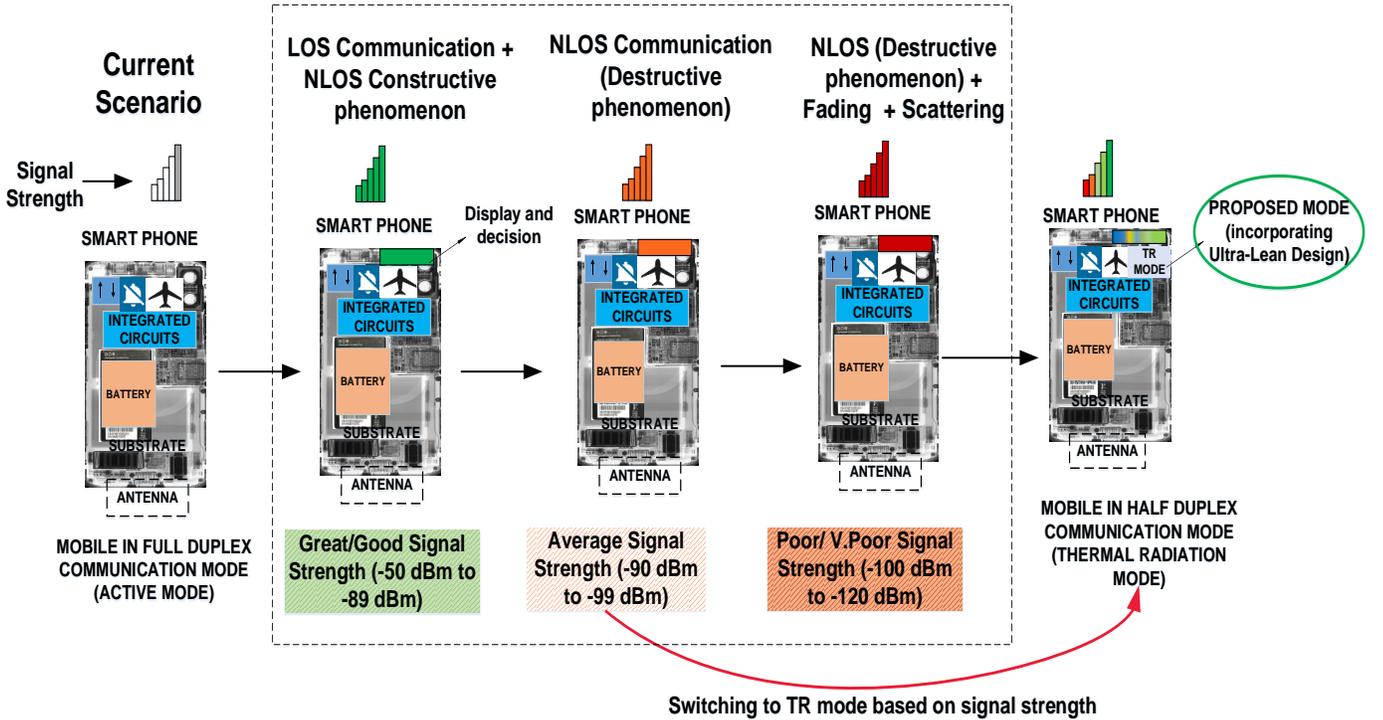

Fig.7 Display and Decision ($D^2$) technique for TR mode.

TABLE III
SIGNAL STRENGTH VALUES FOR DIFFERENT NETWORK
OPERATORS

| Quality of Signal | Jio (4G) | BSNL (Cellone) | Airtel |
|---|---|---|---|
| Great | -75 dBm | -70 dBm | -77dBm |
| Good | -84 dBm | -81 dBm | -85dBm |
| Average | -96 dBm | -91 dBm | -93dBm |
| Poor | -106 dBm | -101 dBm | -104dBm |
| Very Poor | -110 dBm | -105 dBm | -114dBm |

### C. Adaptive Switching Operation from Active Mode (AM) to TR mode

We propose an adaptive switching operation from Active mode (AM) to TR mode for the vehicular users. The adaptive switching takes place when the mobile equipment does not have the appropriate signal strength to support applications requiring high bandwidth. Adaptive switching is based on a training model as depicted in the flowchart shown in Fig. 8. Every user equipment in a cellular network has varying signal strength depending on the different channel conditions with respect to its proximity with the serving base station. Different channel conditions result in the SNR values to be changing correspondingly for every mobile user. When the channel conditions are not favorable the interference power in the network deteriorates the performance of the system hence reducing the maximum achievable capacity or network performance.

The training model prepares a database for every user based on the SNR and SINR of each communicating user. The obtained values for the same are compared with the conditions given in equations 30(a)-30(c). Depending on the SNR, application demand and the target rate $R_t$ to be achieved, decision takes place for a device to remain in Active mode or to transition in TR mode. For a mobile user having signal strength below the threshold value i.e. a signal strength less than the average quality of signal, there is transition to TR mode. When the mobile device transitions into TR mode, the always-on transmission signals are no more active for the device. The uplink information transfer signals are suspended for applications requiring very high bandwidth. This decreases the transmission power of the device as less processing of signals is required.

The reduction in transmission power of the device reduces the power density value which is incident on the human tissue as well as the values for Organ specific averaged SAR and Whole-body averaged SAR. The training model investigates the received SNR at all times so that as soon as the signal strength improves the device can transition into Active mode with all the signals active. This also improves the device battery consumption for the duration the device is in TR mode. Adaptive switching operation prevents the mobile device from excessive heating which results due to operation under poor channel conditions or low signal strength. This aids reduction in the temperature elevation produced in the skin tissues which are particularly vulnerable to radiation induced heating. A device operating in TR mode prevents unsafe overexposure of the human tissues located on the surface of the body which are most sensitive to EM radiation effects.

## V. SIMULATIONS AND DISCUSSIONS

This section presents the analytical results obtained after investigating the biological impact of near-filed exposure due to smart phones. The technical computing is done for mmWave frequencies, as there is a need for modifying the guidelines and regulations for future wireless communication systems especially for devices concerning near-field exposure. The simulation parameters are derived from previous works in literature that study EM radiations or radio frequency radiation impact on human health [10].

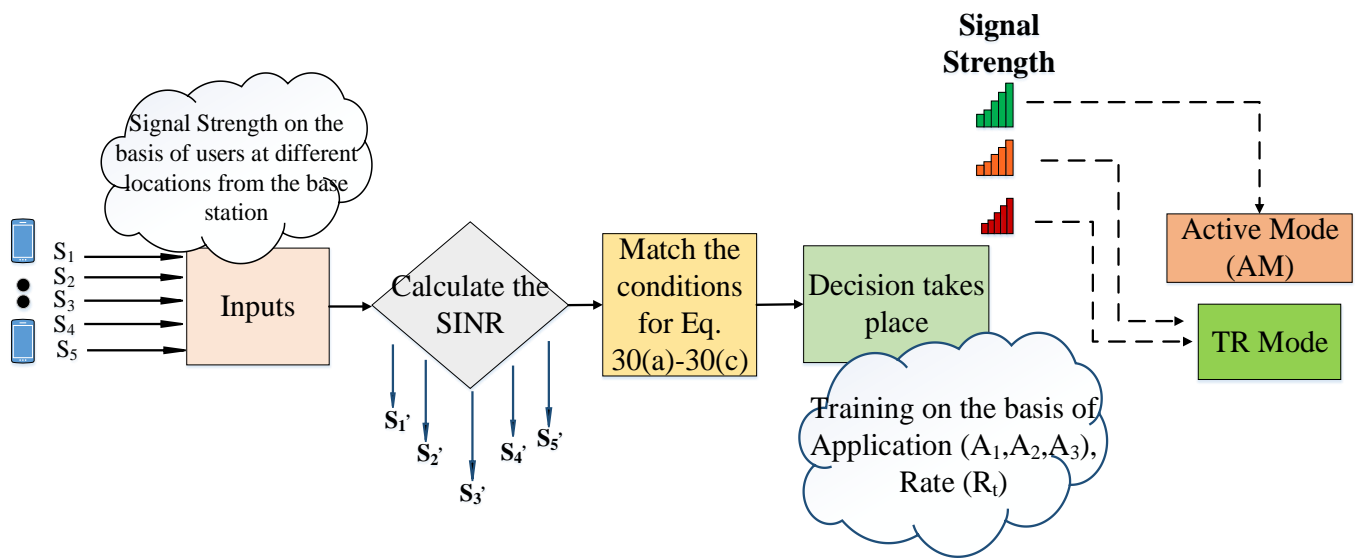

Fig.8 Flowchart for proposed Display and Decision ($D^2$) for TR mode.

The graphical analysis presents a comparison made between the smart phone in "Active mode" and in "Thermal Radiation" (TR) mode for three different applications A3, A2 and A1 as explained in the previous sections. Along with studying the biological impact on human tissues, a comparison has also been made for energy efficiency and complexity of the system for devices operating in the Active and TR mode. To study the power density and temperature rise in the human tissues, numerical computing techniques such as Finite difference time domain (FDTD) has been used. The simulations have been carried out at 30 GHz frequency and the variations in the temperature, power density and SAR have been observed considering the three layer model of the skin. The direct measurements of power density for the 1-D three-layer model of the human tissue have been made which depicts the attenuation of power density with the increasing depth in the skin.

The 1-D three layer model of the skin contains the outermost skin layer (epidermis and dermis) followed by subcutaneous adipose tissue (SAT) and muscle. The power density levels are high in the epidermis and dermis and it gradually decreases with the increase in depth of the tissue. There is very less absorption in the deeper tissue i.e. SAT and muscle. The graph in Fig.9 depicts the comparison of power density levels for the devices operating in "Active mode" and TR mode. As there is maximum power radiated from the device in AM which in turn increases the radiation exposure, the power density levels are highest in the epidermis and dermis layers of the skin tissue. For the devices operating in the TR mode, the radiated power decreases as there is utilization of optimum power to serve the particular application. For the devices operating in TR mode and demanding application A1 i.e. text data, there is least power density absorption in the skin. The significant reduction in power density levels is evident from the graph with the use of TR mode in the device.

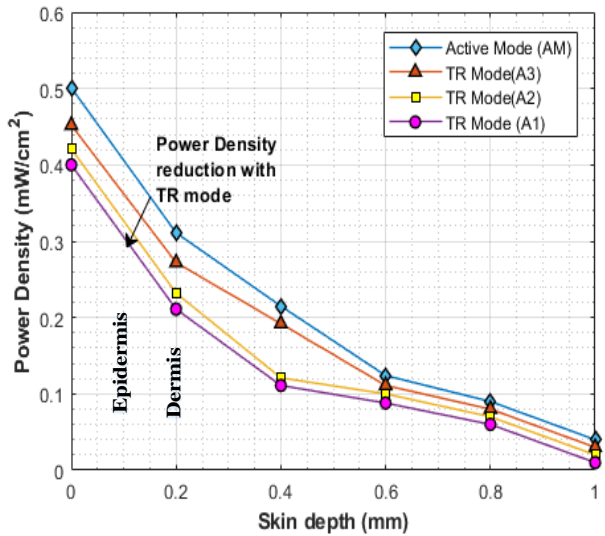

Fig 9. Power Density v/s Skin depth.

Specific absorption rate and its attenuation can be observed from the graph in Fig.10. It is evident that maximum absorption i.e. almost 90% of the transmitted power is absorbed in the epidermis and the dermis layers. There is penetration of power in the deeper layers of the tissues but the absorption levels are quite low. It is apparent from the graph that highest SAR levels are present in the skin for the devices in Active mode and there is an exponential decay as the penetration depth increases. With the incorporation of TR mode, the SAR levels drop subsequently and further decrease with the increase in depth of the tissue.

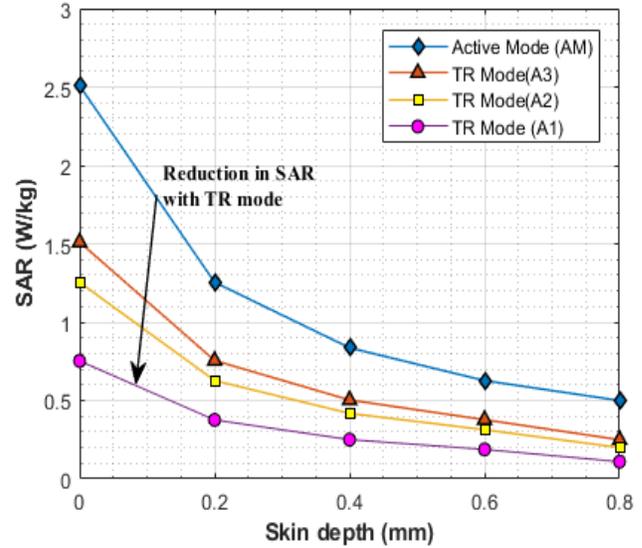

Fig 10. SAR v/s Skin depth.

The devices operating in TR mode consume an optimum value of power required for the particular application and hence decrease the EM radiation levels. As the devices operating in TR mode consume less power, the energy efficiency of the system increases.

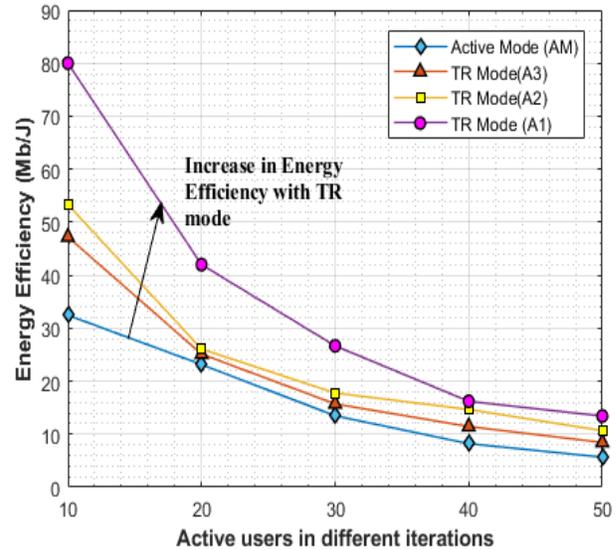

Fig 11. Energy Efficiency v/s Active users in different iterations.

Maximum energy efficiency of the system has been obtained for devices in TR mode with A1 application in function. As the applications become high bandwidth consuming the optimum power requirement also increases substantially and hence the overall energy efficiency of the system decreases. With the TR mode operation in the mobile phone the overall energy efficiency of the system improves considerably as can be seen in Fig.11. Also when the devices are operating in a system, there is interference that occurs due to adjacent channels or co-tier interference. Interference is encountered due to the base station, the cellular devices, devices operating in pairs and so on. Higher the bandwidth consuming application, more is the interference power produced due to its operation. For the devices in TR mode, the utilized power is less and hence the radiated power from the devices is also reduced. This results in an appreciable reduction in complexity of the system with TR mode as is depicted in Fig.12.

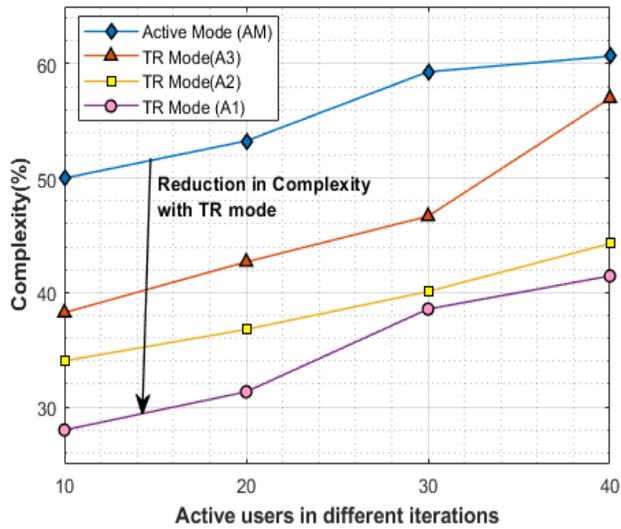

Fig. 12 Complexity v/s Active users in different iterations.

The absorption of non-ionising radiation i.e. the mmWaves in the body is studied with respect to the thermal heating of the tissues. The mmWaves have much smaller wavelengths and hence do not propagate deep into the human tissues. The heating effect is produced due to the temperature elevation that is caused when the electromagnetic radiations are absorbed into the body. There is a rise in temperature that has been observed in the body tissues called as steady-state or transient temperature. If there is overexposure due to mmWaves it can lead to burning sensation on the skin superficially. A temperature elevation of about $0.1^0C$ can be detected by the body as it produces a warm sensation in the skin. A thermal radiation pattern for temperature elevation produced in the human tissue has been depicted in Fig.13.

A comparison has been made for a device operating in Active mode and for a device in TR mode. The 3D radiation pattern in Fig. 13 very clearly illustrates a higher temperature elevation for the device in Active mode and a relatively lower rise in thermal heating for the device operating in TR mode for the same application operating in both the devices. The temperature rise in the tissue is in proportion to the incident power density and the absorption rate in the body. To compute the heat transfer in the human tissue bioheat equation has been used as it incorporates the effect of flow of blood in temperature elevation produced in the tissue.

*Proof*: The proof is given in Appendix B, ***Lemma 2***.  ∎

It is visible from the radiation pattern that the peak of the main lobe has decreased for the device in TR mode and the side lobes correspond to low temperature elevation for a device exhibiting same duration of exposure in both the cases.

## VI. COMPARATIVE ANALYSIS

A comparative analysis has been presented for the proposed TR mode with other popular models in thee literature given by researchers Alekseev and Ziskin [28] and Chahat et al. [29]. They have conducted direct measurements at mmWave frequencies characterizing the complex permittivity of human skin. Debye and Cole-Cole relaxation model have been proposed by them based on the measurement results obtained. The skin dielectric properties and the values of power density attenuation obtained in the skin for different dielectric models are compared with the current communication scenario as well as our proposal. The power density attenuation is obtained for the skin till a depth of 1 mm. The power density values have been used to compute the SAR in the human skin for all the models. The graph obtained after the simulation results is given in Fig.14.

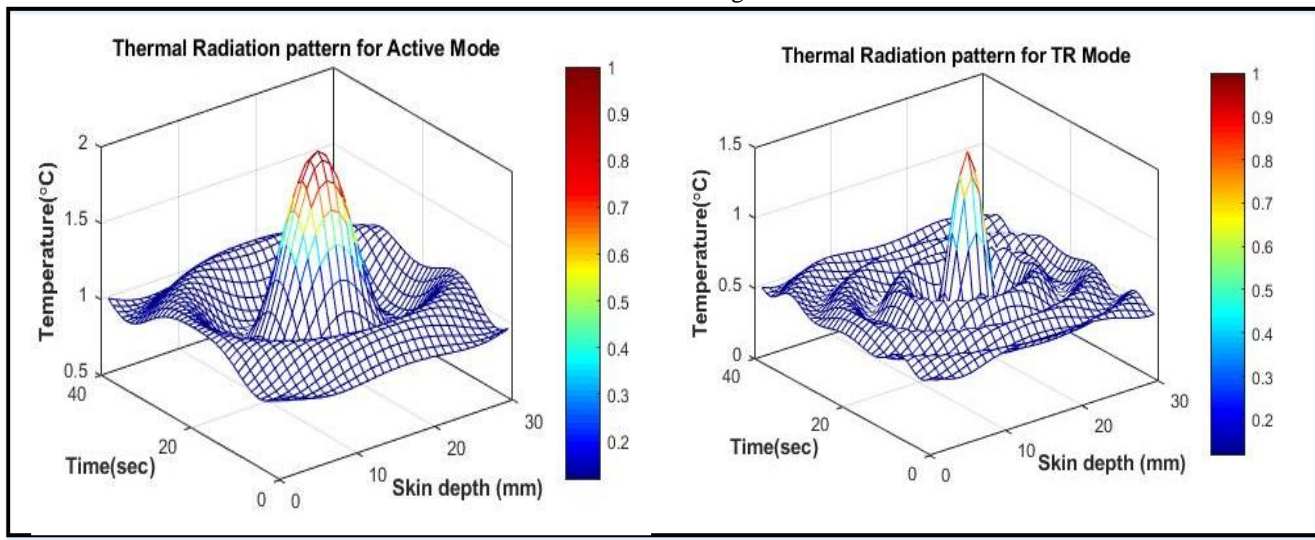

Fig 13. 3D radiation patterns for temperature rise in Active mode and TR mode.

It is evident from the graph that the least values of SAR is obtained for our proposal at varying skin depth due to limitation in the number of signals that are transmitted in the uplink and downlink for a device in TR mode.

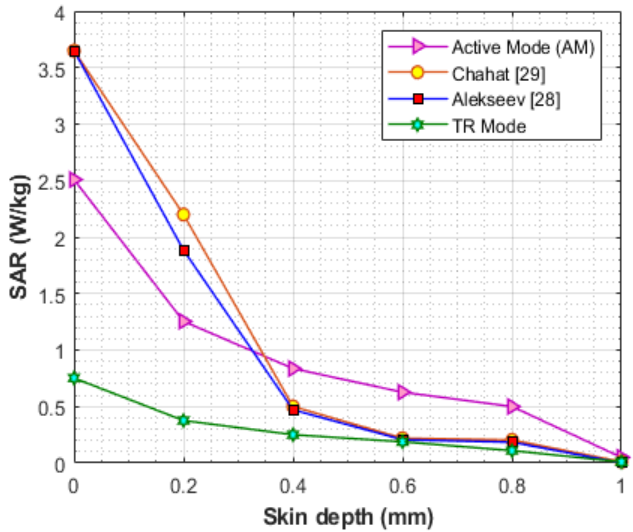

Fig.14 SAR v/s Skin depth.

There is also a reduction in the incident power on the human tissue when the device is communicating in TR mode. A comparative Table IV presents the power density attenuation in the human skin at different depths and percentage (%) reduction in SAR for all the models.

TABLE IV
COMPARATIVE ANALYSIS FOR PD AND SAR

| Power Density ($mW/cm^2$) | | | | | |
|---|---|---|---|---|---|
| Penetration Depth (mm) | Alekseev [31] | Chahat [32] | Active Mode (AM) | (TR) mode | SAR Reduction (%) |
| 0 mm | 0.62 | 0.62 | 0.50 | 0.41 | **Case 1:** w.r.t. Alekseev = 46.75% (↓) |
| 0.2 mm | 0.32 | 0.38 | 0.31 | 0.21 | |
| 0.4 mm | 0.16 | 0.17 | 0.22 | 0.11 | **Case 2:** w.r.t Chahat = 49.88 % (↓) |
| 0.6 mm | 0.09 | 0.10 | 0.14 | 0.09 | |
| 0.8 mm | 0.03 | 0.02 | 0.10 | 0.05 | **Case 3:** w.r.t Active Mode = 69.998 % (↓) |
| 1 mm | 0.01 | 0.01 | 0.05 | 0.01 | |

*The comparative analysis is performed at 0.4mm skin depth.

The comparison for % SAR reduction for all the models is performed at 0.4 mm. The overall reduction in interference power, incident power on the tissue decreases the incident power density and Specific Absorption Rate in the skin for device in TR mode.

## VII. CONCLUSION

This article addresses an escalating area of concern for the future generations of wireless communication networks (WCN).

The impact of rising EM radiation in the environment is highlighted in the article with special emphasis on the biological safety and health hazards associated. Mobile communication systems have been emphasized upon particularly as they contribute majorly to the carbon dioxide footprint and are a source of exposing the body to maximum EM radiation. A proposal in alignment to this has been given in the form of "Thermal radiation" mode i.e. TR mode to limit the EM radiation without hampering the QoS and QoE in the system. The simulation results validate the performance of TR mode in reducing the power density levels, SAR, thermal heating in the body and reduce overall complexity in the system whilst enhancing the energy efficiency. The article insists on the need of a paradigm shift towards making the future generations of communication networks safer, greener and enhancing the existing safety regulations and guidelines in accordance with the current scenario of radiation exposure in the environment.

## APPENDIX A
## PROOF OF LEMMA 1

***Lemma 1:*** The wireless channel considered for the system model is Non-stationary and wideband in nature.

Consider a multipath channel that is time varying in nature, the baseband input signal is denoted as $u(t)$. The received signal at the mobile user is denoted as $r(t)$ which is obtained by convolving the baseband input signal with equivalent low pass time-varying Channel Impulse Response $c(\zeta, t)$ of the channel and then up converting to the carrier frequency (Fig.16).

The signal transmitted from the Access point is represented as:

$$s(t) = \mathcal{R}\{u(t)e^{j(2\pi f_c t + \emptyset_0)}\} \qquad (31)$$

The input signal $u(t) = x(t) + jy(t)$ is a complex baseband signal with in phase component $x(t) = \mathcal{R}\{u(t)\}$, quadrature component $y(t) = \Im\{u(t)\}$, bandwidth B.

The received signal is represented as:

$$r(t) = \mathcal{R}\{v(t)e^{j(2\pi f_c t + \emptyset_0)}\} \qquad (32)$$

The complex baseband signal $v(t)$ depends on the channel through which $s(t)$ propagates.

Here $n = 0$ corresponds to Line of Sight (LOS) path, $f_c$ is the carrier frequency, $\emptyset_0$ is the oscillator phase offset, $\emptyset D_n$ is the Doppler phase shift and $\beta_n(t)$ is the amplitude. $\beta_n(t)$ is a function of path loss and shadowing. $\emptyset_n(t)$ depends on delay, doppler and carrier offset.

The time varying impulse response corresponding to $t_1$ is given as:

$$c(\zeta, t_1) = \sum_{n=0}^{2} \beta_n e^{-j\emptyset_n} \delta(\zeta - \zeta_n) \qquad (33)$$

Time varying impulse response corresponding to $t_2$ equals

$$c(\zeta, t_2) = \sum_{n=0}^{1} \beta'_n e^{-j\emptyset'_n} \delta(\zeta - \zeta'_n) \qquad (34)$$

Table V depicts the numerical values of channel coefficient and channel gain at different time instants for the non-stationary channel considered in our scenario.

The response of the non-stationary channel is depicted in the graph shown Fig. 15. It depicts the energy efficiency corresponding to active users in different iterations for a non-stationary channel for a device in TR mode.

| Time instant | Channel Gain | Channel Coefficient |
|---|---|---|
| t = 0 | g0 =3.319 $\times 10^{-17}$ | 5.76 $\times 10^{-9}$ |
| t = 1 | g1=5.963 $\times 10^{-18}$ | 2.44 $\times 10^{-9}$ |
| t = 2 | g2 = 1.2246 $\times 10^{-18}$ | 1.106 $\times 10^{-9}$ |
| t = 3 | g3=6.692 $\times 10^{-19}$ | 8.18046 $\times 10^{-10}$ |
| t = 4 | g4=3.98 $\times 10^{-19}$ | 6.308 $\times 10^{-10}$ |

At different time instants i.e. at t= 0, 1, 2, 3 the channel coefficients and channel gain do not have large variations and the obtained values of energy efficiency lie in close proximity.

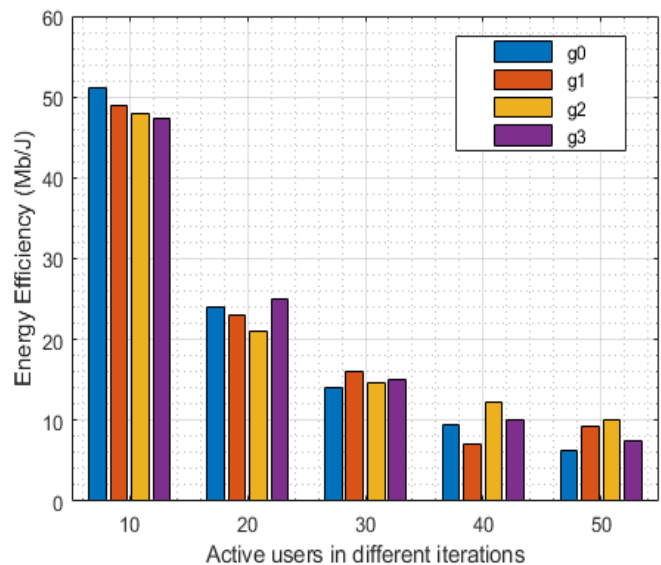

Fig.15 Energy Efficiency v/s Active users in different iterations for Non-stationary Channel.

The different gains corresponding to different time instants can be used to depict the power profile of the non-stationary channel. The plot of the arriving power profile in the fading multipath channel at different instant of time is shown on Fig. 17. It presents the profile of the arriving signal copies. Let us consider $\emptyset(\zeta)$ as the gain and $h(\zeta)$ as the channel coefficient. Then $\emptyset(\zeta) = |h(\zeta)|^2$. The gain can also be written as $\emptyset(\zeta) = \sum_{i=0}^{n} |a_i|^2 \delta(t - \zeta_i)$. Here $|a_i|^2$ is the arriving power. The gain can also be written as $\emptyset(\zeta) = \sum_{i=0}^{n} g_i \, \delta(\zeta - \zeta_i)$. Here $g_i$ is the gain of the ith path.

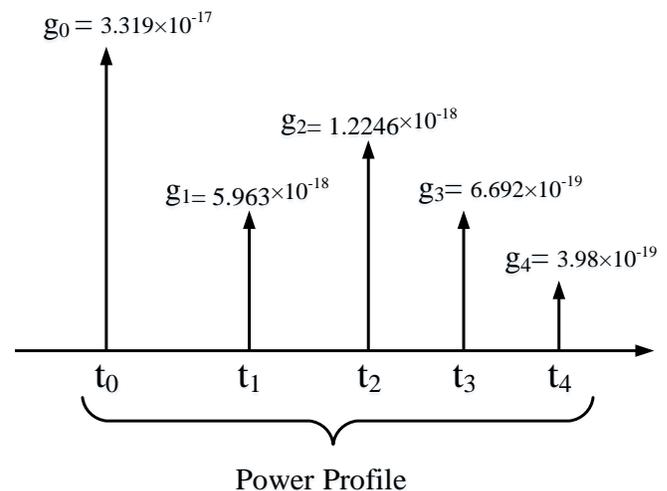

Fig.17 Power delay profile for Non-stationary channel.

## APPENDIX B
## PROOF OF LEMMA 2

*Lemma 2:* The temperature elevation produced in the human tissue for a device operating in TR mode is described using bioheat transfer equation and solved applying finite difference time domain method by expressing it using fourier series. The thermal diffusion equation was modified by Pennes model considering the effects of blood perfusion and metabolism in the human body.

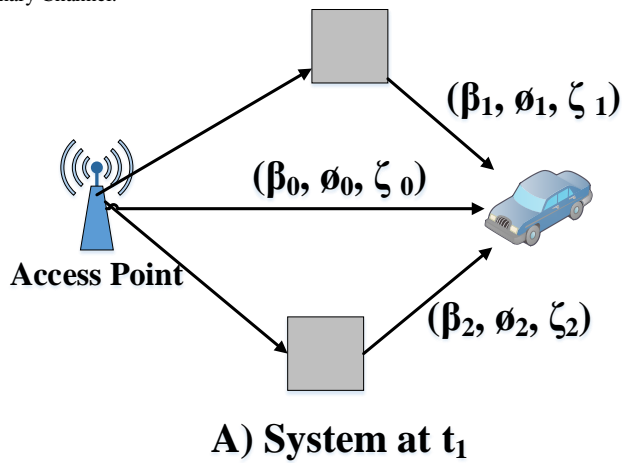

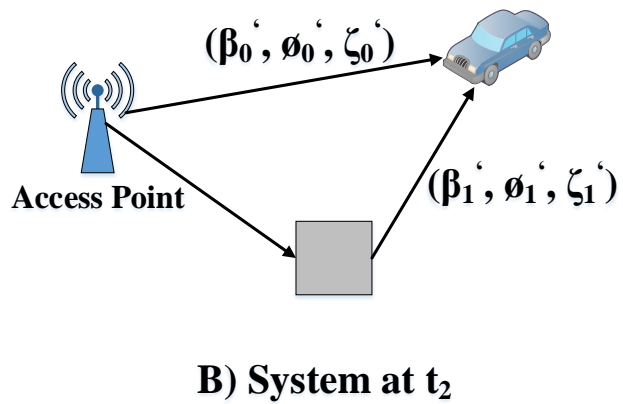

Fig.16 Non-stationary channel with multipath at two different measurement times.

The thermal diffusion equation is as follows

$$\rho_{ts} \, C_{ts} \, \frac{\partial T_{ts}}{\partial t} = \nabla k_{ts} \nabla \, T_{ts} + \rho_b C_b W_b (T_{at} - T_{ts}) + q_h \quad (35)$$

$$q = -k_{ts} \, \nabla T_{ts} + \rho_b \, h_b \, v_b \quad (36)$$

$$h_b = \int_{T_0}^{T_b} C_b \,(T_b) \, dT_b + \frac{P_{sys}}{\rho_b} + \Delta H_m \,(1 - \varphi) \quad (37)$$

Energy balance equation

$$\rho_{ts} \, C_{ts} \, \frac{\partial T_{ts}}{\partial t} = -\nabla . \, q \, , \quad (38)$$

Therefore

$$\rho_{ts} \, C_{ts} \, \frac{\partial T_{ts}}{\partial t} = -\nabla \cdot \left[ -k_{ts} \, \nabla T_{ts} + \rho_b \, v_b \left( \int_{T_0}^{T_b} C_b \,(T_b) dT_b + \frac{P_{sys}}{\rho_b} + \Delta H_f \,(1 - \varphi) \right) \right],$$
$$(39)$$

Simplifying the above equation it can be rewritten as

$$\rho_{ts} \, C_{ts} \, \frac{\partial T_{ts}}{\partial t} = k_{ts} \, \nabla^2 T_{ts} - \rho_b C_b v_b \nabla \, T_b + \rho_b v_b \Delta H_m \nabla \varphi \quad (40)$$

The final form of bio-heat equation is as follows

$$\rho_{ts} \, C_{ts} \, \frac{\partial T_{ts}}{\partial t} = k_{ts} \, \nabla^2 T_{ts} - \rho_b C_b v_b \cdot \nabla T_{ts} + q_h . \quad (41)$$

When we discretize the bio-heat equation using Finite difference time domain method (FDTD) to obtain SAR, the equation obtained is as follows

$$\rho_{ts} \, C_{ts} \, \frac{\partial T}{\partial t} = k_{ts} \, \nabla^2 T_{ts} + \rho \cdot SAR - B_{bl}(T - T_{bl}) \quad (42)$$

With respect to the boundary condition

$$k_{ts} \frac{\partial T}{\partial n} = -h_t (T - T_{am}) \quad (43)$$

where '$n$' is the unit vector normal to the head.
On numerical analysis of the bio-heat equation with FDTD method and expanding it using Fourier series we have

$$\rho_{ts} \, C_{ts} \, \frac{\partial u}{\partial t} = k \nabla^2 u + \rho_{ts} \, SAR - B_{bl} \cdot u \quad (44)$$

As we know that SAR is a ratio of the power dissipated in the tissue to the mass of the tissue that is under exposure. Equation (44) can be re-written as

$$\rho_{ts} \, C_{ts} \, \frac{\partial u}{\partial t} = k \nabla^2 u + \rho_{ts} \frac{P_{in}}{M} - B_{bl} \cdot u \, , \quad (45)$$

For the devices communicating in Active mode i.e. always-on communication signals, the power utilization for a device is computed as $P_{T(max)}^{Total\,(AM)}$. This is the incident power for the exposed body tissues for a user communicating in active mode. Substituting this value of power in equation (45), we get the following equation

$$\rho_{ts} \, C_{ts} \, \frac{\partial u}{\partial t} = k \nabla^2 u + \rho_{ts} \frac{P_{T(max)}^{Total\,(AM)}}{M} - B_{bl} \cdot u \, . \quad (46)$$

The devices communicating in TR mode consume the optimum power $P_{T(opt)}^{Total\,(TR)}$ which is lesser than the power consumed in the active mode. For the devices in TR mode the bio-heat equation can be re-written as follows

$$\rho_{ts} \, C_{ts} \, \frac{\partial u}{\partial t} = k \nabla^2 u + \rho_{ts} \frac{P_{T(opt)}^{Total\,(TR)}}{M} - B_{bl} \cdot u \, , \quad (47)$$

The finite difference approximation of the above equation can be written as follows

$$u^{i+1}(x,y,z) = u^i(x,y,z) - $$
$$\frac{\delta_t}{\rho(x,y,z)C_p(x,y,z)} b(x,y,z)u^i(x,y,z) + \frac{\delta_t \, k_{ts}(x,y,z)}{\rho(x,y,z)C_p(x,y,z)\delta^2} \cdot$$
$$[u^i(x+1,y,z) + u^i(x,y+1,z) + u^i(x,y,z+1) +$$
$$u^i(x-1,y,z) + u^i(x,y-1,z) + u^i(x,y,z-1) -$$
$$6u^i(x,y,z) \, , \quad (48)$$

where $i$ is the iteration number. On expansion of '$u$' utilizing fourier series, we have

$$u(x,y,z,i) = \sum_{\lambda=0}^{x_0-1} \sum_{\mu=0}^{y_0-1} \sum_{\gamma=0}^{z_0-1} U_{\lambda,\mu,\gamma}^i \cdot \exp[\, \aleph(x \, \frac{2\pi}{x_0} \lambda + y \, \frac{2\pi}{y_0} \mu + z \, \frac{2\pi}{z_0} \gamma)] \, , \quad (49)$$

$$U_{\lambda,\mu,\gamma}^i = \frac{1}{x_0 y_0 z_0} \sum_{x=0}^{x_0-1} \sum_{y=0}^{y_0-1} \sum_{z=0}^{z_0-1} u(x,y,z,i) \cdot \exp[-\aleph(\lambda \frac{2\pi}{x_0} x + \mu \frac{2\pi}{y_0} y + \gamma \frac{2\pi}{z_0} z)] \, , \quad (50)$$

where $u(x,y,z,i)$ is assumed within the range as $\{x\} = \{0,1,2, \ldots \ldots x_0 - 1\}, \{y\} = \{0,1,2, \ldots \ldots y_0 - 1\}, \{z\} = \{0,1,2, \ldots \ldots z_0 - 1\}, \{i\} = \{0,1,2, \ldots \ldots \binom{T_t}{\delta_t}\}$ and $\aleph = \sqrt{-1}$
Substituting (50) in (48) and dividing by $e^{\aleph(x(2\pi/x_0)\,\lambda + y(\frac{2\pi}{y_0})\mu + z(\frac{2\pi}{z_0})\gamma)}$, one fourier component can be expressed as

$$U_{\lambda,\mu,\gamma}^{i+1} = [1 - \frac{\delta_t \, b}{\rho_{ts} \, C_{ts}} - \frac{4 \, \delta_t k}{\rho_{ts} \, C_{ts} \, \delta^2} \, \{Sin^2 \left( \frac{\pi}{x_0} \lambda + Sin^2 \left( \frac{\pi}{y_0} \mu + Sin^2 \left( \frac{\pi}{z_0} \gamma \right) \right\}] \cdot U_{\lambda,\mu,\gamma}^i \, . \quad (51)$$

Considering '$w$' as an incremental rate for each of the fourier component at time step $\delta_t$ we can write $U_{\lambda,\mu,\gamma}^{i+1} = w \, U_{\lambda,\mu,\gamma}^i$
For obtaining a stable solution for all values of $i$, $\lambda$, $\mu$ and $\gamma$, we consider an integer $N$ which fulfills the following condition

$$w \leq 1 + N\delta_t \quad (52)$$

The above equation (52) is known as Von Neumann's condition [30]. On utilizing this condition in equation (51) we get
The integer $N$ must always fulfill the following condition for all values of $\lambda$, $\mu$ and $\gamma$

$$\delta_t \leq \frac{2 \, \delta_{ts} \, C_{ts} \, \delta^2}{12k + b\delta^2} \quad (53)$$

## ACKNOWLEDGEMENT

The authors gratefully acknowledge the support provided by 5G and IoT Lab, SoECE, TBIC, TEQIP-III at Shri Mata Vaishno Devi University, Katra, Jammu. This work has been patented with the title "Thermal Radiation (TR) mode for Electromagnetic radiation reduction in future wireless networks".

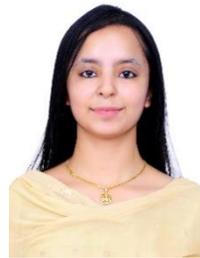

**HANEET KOUR** (S'17) received the B.E degree in ECE engineering from Jammu University, Jammu and Kashmir, India in 2015 and the M.Tech degree in ECE Engineering from Shri Mata Vaishno Devi University in 2017. She is currently pursuing the Ph. D degree in Electronics and Communication Engineering at Shri Mata Vaishno Devi University, Katra, Jammu and Kashmir, India. Her research interest includes the emerging technologies of 5G wireless communication network. Currently she is doing her research on Power Optimization in next generation networks. She is working on MATLAB tools for Wireless Communication. She is a student member of Institute of Electrical and Electronics Engineers (IEEE).

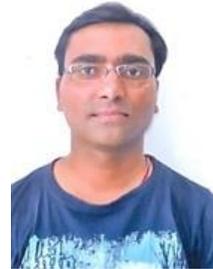

**Dr. Rakesh K Jha (S'10, M'13, SM 2015)** is currently an Associate Professor in School of ECE Engineering, Shri Mata Vaishno Devi University, Katra, Jammu and Kashmir, India. He is carrying out his research in wireless communication, power optimizations, wireless security issues and optical communications. He has done B.Tech. in ECE Engineering from Bhopal, India and M. Tech from NIT Jalandhar, India. He received his PhD degree from NIT Surat, India in 2013.

He has published more than 41 Science Citation Index Journals Papers including many IEEE Transactions, IEEE Journal and more than 25 International Conference papers. His area of interest is Wireless communication, Optical Fiber Communication, Computer networks, and Security issues. Dr. Jha's one concept related to router of Wireless Communication has been accepted by ITU in 2010. He has received young scientist author award by ITU in Dec 2010. He has received APAN fellowship in 2011, 2012, 2017 and 2018 and student travel grant from COMSNET 2012. He is a senior member of IEEE, GISFI and SIAM, IAENG and ACCS. He is also member of ACM and CSI, many patents and more than 2800 Citations in his credit.